\documentclass[11pt]{article}

\usepackage[final]{acl}

\usepackage{times}
\usepackage{latexsym}

\usepackage[T1]{fontenc}

\usepackage[utf8]{inputenc}

\usepackage{microtype}

\usepackage{inconsolata}

\usepackage{graphicx}
\usepackage{subcaption}
\usepackage{multirow}

\usepackage{algorithm}
\usepackage{algpseudocode}

\usepackage{amsmath}
\usepackage{amssymb}
\usepackage{mathtools}
\usepackage[table]{xcolor}
\usepackage[most]{tcolorbox}
\usepackage{xspace}

\usepackage[most]{tcolorbox}

\newtcolorbox{fedagentbox}[2][]{
  breakable,
  colback=gray!3,
  colframe=blue!45!black,
  colbacktitle=blue!45!black,
  coltitle=white,
  fonttitle=\bfseries,
  title={#2},
  boxrule=0.45pt,
  arc=1mm,
  left=1mm,
  right=1mm,
  top=1mm,
  bottom=1mm,
  #1
}

\title{FedAgentKE: Federated Semantic Knowledge Evolution for Heterogeneous Agents}

\author{
Weihao Li$^{1*}$ \quad
Jun Bai$^{2,3*}$ \quad
Ziyang Song$^{4\dagger}$ \\[2mm]
$^{1}$McCormick School of Engineering, Northwestern University \\
$^{2}$School of Computer Science, McGill University \\
$^{3}$Mila -- Quebec AI Institute \\
$^{4}$School of Electrical Engineering and Computer Science, Ohio University \\[1mm]
\texttt{weihaoli2027@u.northwestern.edu},
\texttt{jun.bai@mcgill.ca},
\texttt{ziyangs@ohio.edu}
}

\begin{document}

\maketitle

\begingroup
\renewcommand{\thefootnote}{}
\footnotetext{\hspace{0.0em}$^{*}$ Equal contribution.}
\footnotetext{\hspace{0.0em}$^{\dagger}$ Corresponding author.}
\addtocounter{footnote}{-2}
\endgroup

\begin{abstract}
Large language model (LLM)-based agents increasingly rely on reasoning, tool use, and iterative execution, yet existing agent frameworks still operate largely in isolation. While recent memory-based agent systems improve individual agents through local retrieval and workflow reuse, local experiences remain fragmented across isolated agent frameworks, limiting cross-framework knowledge transfer and collaborative reasoning evolution. We propose \textbf{FedAgentKE}, a lightweight framework for \textbf{Federated Semantic Knowledge Evolution} across heterogeneous agents. 
FedAgentKE enables distributed agent frameworks to collaboratively evolve transferable reasoning abstractions through iterative semantic knowledge distillation, aggregation, and adaptation without sharing raw reasoning trajectories.
Experiments demonstrate consistent improvements under both cross-framework and cross-task settings, highlighting the potential of federated semantic knowledge evolution for future collaborative agent ecosystems.

\end{abstract}


\section{Introduction}
Large language model (LLM)-based agents are increasingly used to solve complex tasks through multi-step reasoning, tool use, and iterative execution \citep{yao2022react, hong2024metagpt, qin2024toolllm}. Recent agent frameworks, such as OpenHands \citep{wang2025openhands}, OWL \citep{hu2026owl}, and SmolAgents \citep{smolagents}, have demonstrated strong capabilities across software engineering, web interaction, and long-horizon reasoning tasks. 
However, existing agent frameworks still learn largely in isolation, where reasoning experiences and execution feedback remain fragmented within individual agent ecosystems. 
As a result, heterogeneous agents often rediscover similar solutions and repeat avoidable failures, limiting collaborative knowledge evolution across agent frameworks.


To improve agent reasoning and planning, prior work has explored memory-augmented and knowledge-enhanced agent systems \citep{shinn2023reflexion, wang2024agent, xu2026mem, zhu2025knowagent}. These methods reuse past trajectories, retrieve  external memories, or refine workflows. For instance, Reflexion \citep{shinn2023reflexion} improves agents through verbal self-feedback, while A-MEM \citep{xu2026mem} builds agentic memory for experience reuse. More recently, Agent KB \citep{tang2025agentkb} enables cross-framework experience sharing through a centralized knowledge base.
Nevertheless, existing memory-based approaches still mainly rely on local memory reuse or centralized knowledge storage, providing limited support for cross-framework knowledge synchronization and adaptation. This leaves \textbf{\textit{collaborative knowledge evolution across heterogeneous agent ecosystems largely unexplored}}.

This gap raises a key question: \textit{how can heterogeneous agents collaboratively improve without sharing raw execution trajectories?} 
Such decentralized agent ecosystems naturally motivate collaboration in a federated learning (FL) manner \citep{mcmahan2017communication, bai2025unified}, where distributed agents can exchange transferable knowledge without exposing raw execution experiences.
However, directly applying traditional FL to modern agent systems remains challenging, since heterogeneous agent frameworks often differ substantially in reasoning protocols, tool environments, and execution workflows~\cite{xu2026rethinking}. Moreover, many transferable capabilities in agent systems are not encoded in neural parameters, but instead emerge as reusable reasoning abstractions and execution strategies.
This motivates federated semantic knowledge evolution across agents.

In this study, we propose \textbf{FedAgentKE}, a lightweight framework for \textbf{Federated Semantic Knowledge Evolution} across heterogeneous agents. FedAgentKE distills local agents' execution experiences into transferable semantic knowledge units and synchronizes them through a federated knowledge pool. By operating at the semantic level rather than the parameter level, FedAgentKE enables collaborative reasoning improvement across heterogeneous agent frameworks without sharing full execution trajectories. We evaluate FedAgentKE across diverse agent frameworks and tasks under decentralized settings. Experimental results show that FedAgentKE consistently improves performance under both cross-framework and cross-task settings. Performance further improves as more heterogeneous agents participate in federated synchronization, suggesting the scalability of semantic-level knowledge evolution for collaborative agent ecosystems.

Our main contributions are threefold: (1) We formulate heterogeneous agent collaboration as a new \textbf{federated semantic knowledge evolution} problem, enabling collaborative reasoning evolution without sharing raw execution trajectories. (2) We propose \textbf{FedAgentKE}, a lightweight framework that enables semantic-level knowledge synchronization across heterogeneous agent frameworks through transferable semantic knowledge distillation and federated aggregation. (3) We evaluate FedAgentKE across multiple heterogeneous agent frameworks and tasks, demonstrating effective cross-framework knowledge transfer and collaborative knowledge evolution in decentralized agent ecosystems.







\section{FedAgentKE} \label{sec:method}

\subsection{Framework Overview}

\begin{figure*}[t]
\centering
\includegraphics[width=0.9\textwidth]{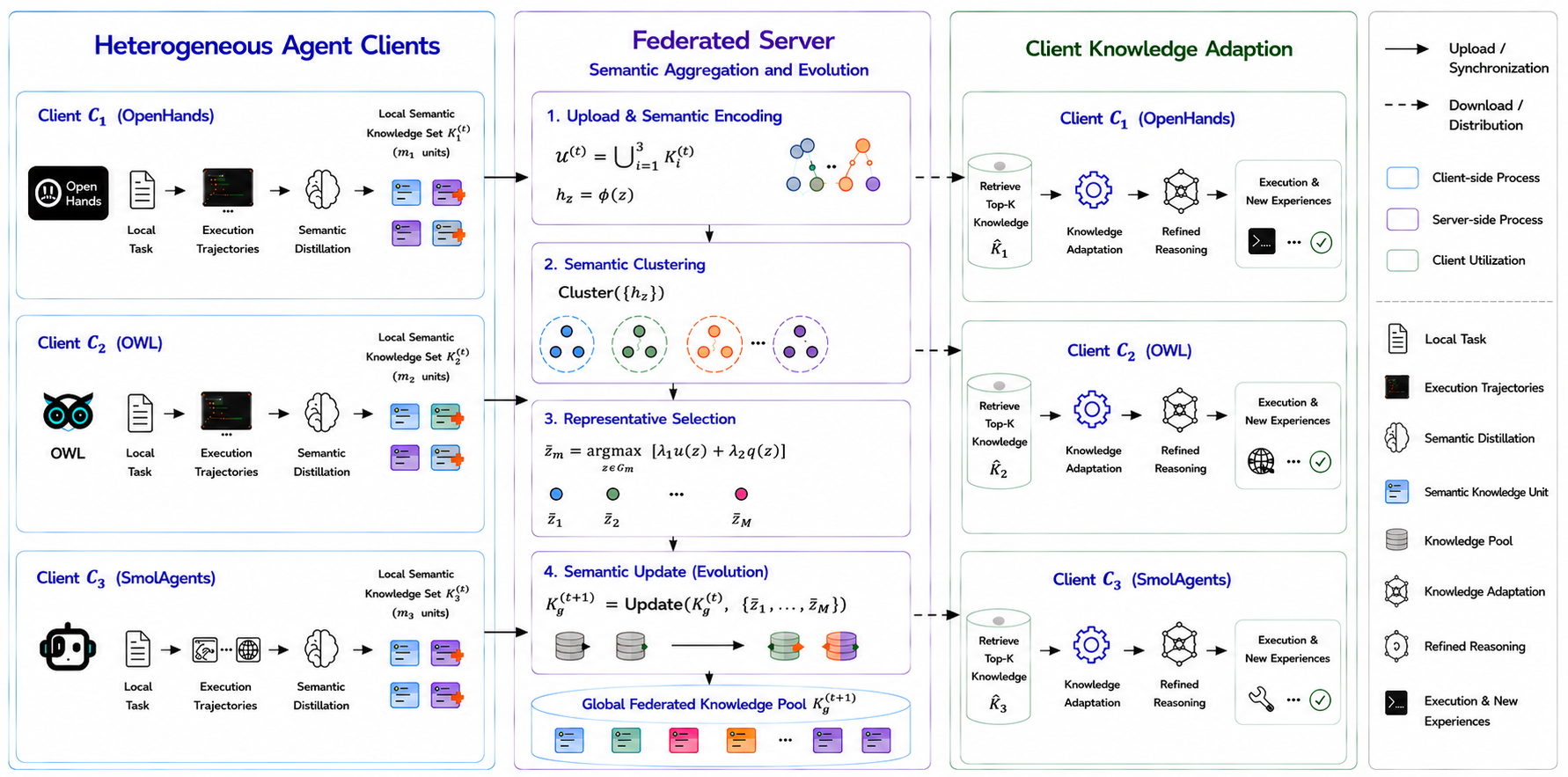}
\vspace{-2mm}
\caption{\small
Overview of FedAgentKE. Heterogeneous agents collaboratively evolve through federated semantic knowledge synchronization across diverse frameworks and tasks.
}
\label{fig:framework}
\vspace{-4mm}
\end{figure*}


We formulate heterogeneous agent collaboration as a federated semantic knowledge evolution problem, where distributed agent clients collaboratively improve through synchronization of semantic knowledge  distilled from local execution experiences.

Consider a set of $N$ heterogeneous agent clients,
$
\mathcal{C}=\{C_1,C_2,\ldots,C_N\}.
$
At communication round $t$, each client $C_i$ maintains a local knowledge set:
\begin{equation}
K_i^{(t)}
=
\{z_{i,1}^{(t)},z_{i,2}^{(t)},\ldots,z_{i,m_i}^{(t)}\},
\end{equation}
where $m_i$ denotes the number of local semantic knowledge units maintained by client $C_i$, and each semantic knowledge unit $z_{i,j}^{(t)}$ represents a transferable reasoning abstraction distilled from local experiences. The server maintains a global knowledge pool $K_g^{(t)}$ for federated semantic synchronization.


The overall framework is illustrated in Figure~\ref{fig:framework}. FedAgentKE consists of three stages: local knowledge distillation, federated semantic aggregation, and knowledge adaptation. Each agent distills execution experiences into transferable semantic knowledge units, which are aggregated at the federated server and redistributed for local  adaptation. Through iterative synchronization and utility-aware semantic updating, the federated knowledge pool continuously evolves across agent ecosystems. The complete algorithm is provided in Appendix~\ref{app:algorithm}.

\subsection{Local Knowledge Distillation}


Heterogeneous agent frameworks often generate framework-specific execution trajectories, making direct cross-agent reuse difficult. For example, OpenHands may produce repository-level code-editing trajectories, while SmolAgents may generate web-interaction and tool-calling workflows. FedAgentKE therefore abstracts local execution trajectories into transferable semantic knowledge representations. Specifically, let
\begin{equation}
\tau_{i,j}
=
(x_{i,j},a_{i,j}^{1:T},o_{i,j}^{1:T},y_{i,j})
\end{equation}
denote the $j$-th execution trajectory collected from client $C_i$, where $x_{i,j}$ is the task input, $a_{i,j}^{1:T}$ and $o_{i,j}^{1:T}$ denote the action and observation sequences over $T$ execution steps, respectively, and $y_{i,j}$ denotes the final execution outcome.

FedAgentKE then applies a local semantic distillation operator:
\begin{equation}
z_{i,j}
=
\mathcal{D}_i(\tau_{i,j}),
\end{equation}
where $\mathcal D_i(\cdot)$ denotes an LLM-based semantic abstraction module. The module converts framework-specific trajectories into transferable semantic knowledge units by summarizing reusable reasoning patterns, execution corrections, and workflow structures while filtering framework-specific implementation details. 
Each distilled semantic knowledge unit is represented as:
\begin{equation}
z_{i,j}
=
(\pi_{i,j},\rho_{i,j},e_{i,j},c_{i,j},u_{i,j}),
\end{equation}
where $\pi_{i,j}$ denotes the abstracted task description, $\rho_{i,j}$ represents reusable reasoning workflows, $e_{i,j}$ stores execution corrections or failure patterns, $c_{i,j}$ contains framework metadata, and $u_{i,j}$ denotes the utility score of the knowledge unit.

Semantic distillation preserves transferable reasoning patterns and execution experiences while reducing framework-specific dependence.

\subsection{Federated Semantic Aggregation}

After local distillation, clients upload knowledge units to the federated server for collaborative aggregation. A naive union of all uploaded knowledge may introduce redundancy, conflicting reasoning patterns, and low-quality experiences, particularly under heterogeneous tasks and agent environments. FedAgentKE therefore performs task-aware semantic aggregation rather than direct memory merging

We denote the uploaded knowledge pool at communication round $t$ as:
\begin{equation}
\mathcal{U}^{(t)}
=
\bigcup_{i=1}^{N} K_i^{(t)}.
\end{equation}
Each semantic knowledge unit $z$ is first encoded into a semantic representation:
\begin{equation}
h_z=\phi(z),
\end{equation}
where $\phi(\cdot)$ denotes a semantic embedding function.

FedAgentKE then groups related knowledge units through cross-task semantic clustering:
\begin{equation}
\mathcal{G}_1,\ldots,\mathcal{G}_M
=
\operatorname{Cluster}
\left(
\{h_z \mid z \in \mathcal{U}^{(t)}\}
\right),
\end{equation}
where semantically related reasoning abstractions with cross-task transferability are grouped together.

For each cluster $\mathcal{G}_m$, FedAgentKE selects representative semantic knowledge according to both empirical utility and cross-agent transferability:
\begin{equation}
\bar{z}_m
=
\arg\max_{z \in \mathcal{G}_m}
\left[
\lambda_1 u(z)
+
\lambda_2 q(z)
\right],
\end{equation}
where $u(z)$ measures the empirical usefulness of knowledge unit $z$, and $q(z)$ measures its transferability across heterogeneous agents and tasks. A utility-based interpretation of cross-agent knowledge transferability is provided in Appendix~\ref{app:theory}.

The selected representative knowledge units are then integrated into the global knowledge pool:
\begin{equation}
K_g^{(t+1)}
=
\operatorname{Update}
\left(
K_g^{(t)},
\{\bar{z}_1,\ldots,\bar{z}_M\}
\right),
\end{equation}
where $\operatorname{Update}(\cdot)$ performs semantic deduplication and representative knowledge updating. Detailed semantic update and knowledge replacement mechanisms are provided in Appendix~\ref{app:semantic_update}.

\subsection{Knowledge Adaptation}

The aggregated federated knowledge is redistributed to clients for knowledge-guided reasoning. Given a local task $x_i$, client $C_i$ retrieves the top-$k$ semantically relevant federated knowledge units:
\begin{equation}
\hat{K}_i
=
\operatorname{TopK}_{z \in K_g}
\operatorname{sim}
\left(
\psi(x_i),\phi(z)
\right),
\end{equation}
where $\psi(x_i)$ encodes the task context, $\phi(z)$ denotes the embedding of knowledge unit $z$, and $\operatorname{sim}(\cdot,\cdot)$ measures similarity in the shared embedding space.

Since directly applying external reasoning abstractions may introduce incompatibility, FedAgentKE performs knowledge adaptation:
\begin{equation}
\tilde{K}_i
=
\mathcal{T}_i(\hat{K}_i,c_i),
\end{equation}
where $c_i$ denotes the local framework context and $\mathcal{T}_i(\cdot)$ maps federated knowledge into locally executable reasoning guidance. The adapted knowledge is then used to refine local reasoning:
\begin{equation}
\rho_i'
=
\mathcal{R}_i
\left(
\rho_i,\tilde{K}_i
\right),
\end{equation}
where $\mathcal{R}_i(\cdot)$ denotes the local reasoning refinement operator, and $\rho_i$ and $\rho_i'$ denote the original and refined reasoning workflows, respectively.





\section{Experiments}

\subsection{Experimental Setup}


We evaluate \textbf{FedAgentKE} on general reasoning and software engineering tasks using \textbf{GAIA}~\citep{mialon2024gaia} and \textbf{SWE-bench Lite}~\citep{jimenez2024swe}. GAIA contains 53 Level-1, 86 Level-2, and 26 Level-3 tasks for multi-step reasoning and tool use. SWE-bench Lite evaluates real-world GitHub issue-resolution tasks, where we use 300 sampled cases following the standard protocol.We report task success rate on both benchmarks.



We consider four representative agent frameworks, including \textbf{OWL}, \textbf{SmolAgents}, \textbf{OpenHands}, and \textbf{SWE-agent}~\citep{yang2024swe}. Unless otherwise specified, all experiments use \textbf{GPT-5.4-mini} as the backbone model. FedAgentKE uses 3 distributed agent clients, where local tasks are randomly partitioned without overlap across clients under balanced workload constraints. For SWE-bench Lite experiments, each local agent uses a maximum execution iteration budget of 50.


\subsection{Results}

\begin{table}[t]
\centering
\small
\fontsize{8}{9.2}\selectfont
\setlength{\tabcolsep}{1pt}
\captionsetup{skip=5pt}
\caption{ \small
Intra-framework agent federation results. $R$ denotes the number of federated rounds. GAIA (Avg) represents the average performance across Levels 1--3 (L1--3). Standalone denotes the original agent without federated knowledge synchronization. Improvements are computed over Standalone.
}
\begin{tabular}{ccccc}
\hline
\multirow{2}{*}{\textbf{\scriptsize Benchmark}} &
\multirow{2}{*}{\textbf{\scriptsize Framework}} &
\multirow{2}{*}{\textbf{\scriptsize Standalone}} &
\multicolumn{2}{c}{\textbf{\scriptsize FedAgentKE}} \\
\cline{4-5}
 & & & \textbf{$R=1$} & \textbf{$R=2$} \\
\hline

\multirow{2}{*}{\shortstack{\scriptsize SWE-bench \\ Lite}}
& SWE-agent
& 27.00
& 32.33 {\scriptsize($\uparrow$5.33)}
& 42.67 {\scriptsize($\uparrow$15.67)} \\

& OpenHands
& 29.33
& 36.00 {\scriptsize($\uparrow$6.67)}
& 44.67 {\scriptsize($\uparrow$15.34)} \\

\hline

\multirow{2}{*}{GAIA (Avg)}
& OWL
& 40.61
& 46.06 {\scriptsize($\uparrow$5.45)}
& 60.00 {\scriptsize($\uparrow$19.39)} \\

& SmolAgents
& 41.21
& 44.24 {\scriptsize($\uparrow$3.03)}
& 54.55 {\scriptsize($\uparrow$13.34)} \\

\hline

\multirow{2}{*}{GAIA L1}
& OWL
& 56.60
& 75.47 {\scriptsize($\uparrow$18.87)}
& 79.25 {\scriptsize($\uparrow$22.65)} \\

& SmolAgents
& 50.94
& 62.26 {\scriptsize($\uparrow$11.32)}
& 75.47 {\scriptsize($\uparrow$24.53)} \\

\hline

\multirow{2}{*}{GAIA L2}
& OWL
& 36.05
& 44.19 {\scriptsize($\uparrow$8.14)}
& 52.33 {\scriptsize($\uparrow$16.28)} \\

& SmolAgents
& 39.53
& 46.51 {\scriptsize($\uparrow$6.98)}
& 52.33 {\scriptsize($\uparrow$12.80)} \\

\hline

\multirow{2}{*}{GAIA L3}
& OWL
& 23.08
& 30.77 {\scriptsize($\uparrow$7.69)}
& 42.31 {\scriptsize($\uparrow$19.23)} \\

& SmolAgents
& 26.92
& 34.62 {\scriptsize($\uparrow$7.70)}
& 42.31 {\scriptsize($\uparrow$15.39)} \\

\hline
\end{tabular}
\label{tab:intra_framework}
\end{table}

\paragraph{Intra-framework Agent Federation}
We first evaluate whether federated semantic knowledge evolution improves agents within the same framework under different local tasks. Compared with standalone agents, FedAgentKE consistently improves frameworks on both SWE-bench Lite and GAIA. On SWE-bench Lite, SWE-agent improves from 27.00 to 42.67, while OpenHands improves from 29.33 to 44.67 after two federated rounds. On GAIA, the average gains over standalone agents reach 19.39 points for OWL and 13.34 for SmolAgents, demonstrating effective knowledge transfer within the same agent framework.

\paragraph{Cross-framework Agent Federation}

\begin{figure}[t]
\centering
\captionsetup{skip=5pt}
\includegraphics[width=0.98\linewidth]{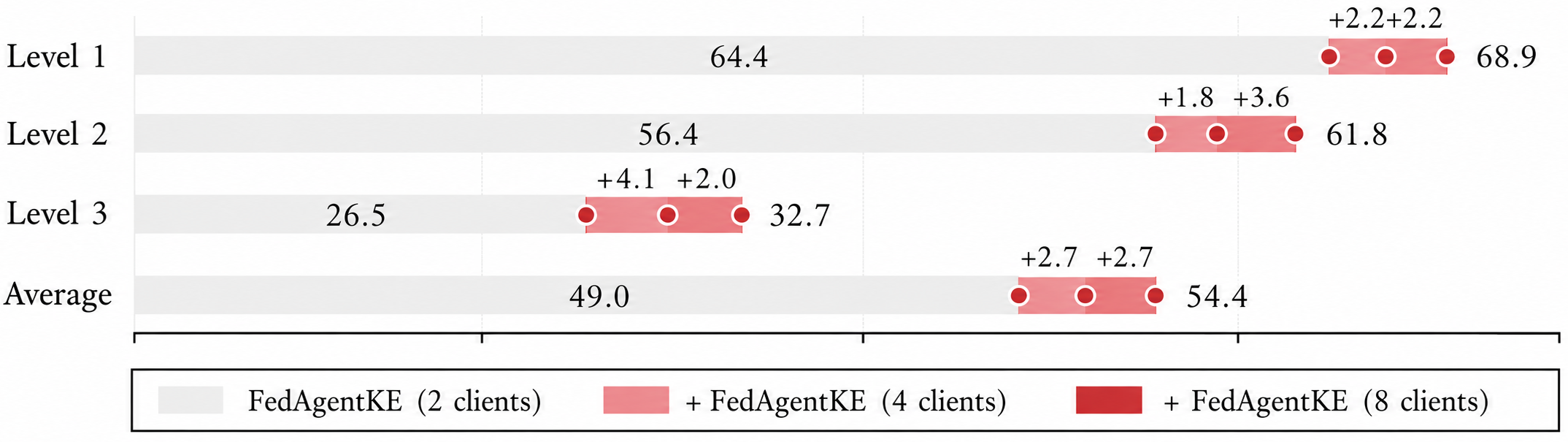}
\caption{\small
Cross-framework agent federation results on GAIA with varying client scales. Clients are randomly instantiated from OWL and SmolAgents. All larger federations are evaluated as incremental extensions of the original 2-client federation consisting of one OWL client and one SmolAgents client.
}
\label{fig:cross-agent}
\end{figure}

We evaluate cross-framework agent federation under different client scales on GAIA. Agent clients are randomly instantiated from OWL and SmolAgents. Figure~\ref{fig:cross-agent} shows that increasing the number of heterogeneous clients consistently improves federated performance across all GAIA difficulty levels, suggesting that transferable semantic reasoning experiences can be effectively synchronized across heterogeneous agent frameworks. The overall average success rate improves from 49.0\% under 2 clients to 54.4\% under 8 clients, indicating that larger heterogeneous federations provide richer collaborative knowledge for semantic knowledge evolution.

\paragraph{Evolution across Rounds}

\begin{figure}[t]
\centering
\captionsetup{skip=5pt}
\includegraphics[width=0.98\linewidth]{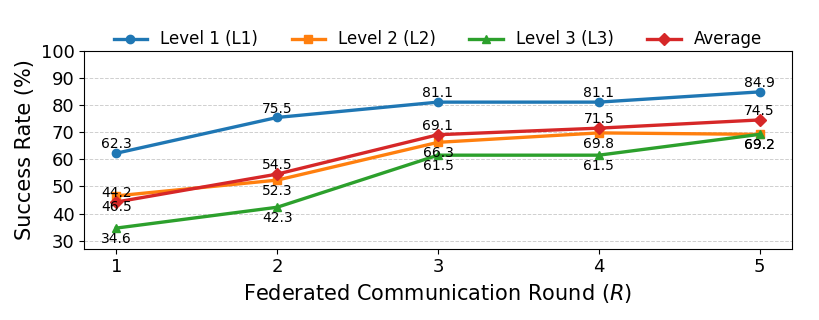}
\caption{\small
Performance evolution across federated communication rounds on GAIA using SmolAgents.
}
\label{fig:round_evolution}
\end{figure}


We further evaluate how iterative federated knowledge evolution affects performance across communication rounds using SmolAgents on GAIA. As shown in Figure~\ref{fig:round_evolution}, performance consistently improves as rounds increase. The overall success rate improves from 44.24\% at $R=1$ to 74.55\% at $R=5$, while similar trends are observed across all GAIA levels. These results suggest that iterative federated knowledge synchronization progressively improves transferable reasoning and execution capabilities across rounds.



\section{Conclusion}




FedAgentKE presents an initial exploration of federated semantic knowledge evolution for heterogeneous agents, showing that distributed agent systems can collaboratively improve without sharing raw execution trajectories. 

\section*{Limitations}


FedAgentKE currently focuses on semantic knowledge synchronization across heterogeneous agents, without explicitly modeling communication efficiency or privacy-preserving knowledge sharing. In addition, the current embedding-based semantic update mechanism may not fully capture complex reasoning dependencies across diverse agent environments. Finally, our experiments mainly evaluate several representative agent frameworks and benchmarks, while evaluations on larger-scale heterogeneous agent ecosystems remain future work.




\bibliography{custom}

\begin{thebibliography}{17}
\providecommand{\natexlab}[1]{#1}

\bibitem[{Bai et~al.(2025)Bai, Song, Wu, Sajjanhar, Xiang, Zhou, Tao, Li, and Li}]{bai2025unified}
Jun Bai, Yiliao Song, Di~Wu, Atul Sajjanhar, Yong Xiang, Wei Zhou, Xiaohui Tao, Yan Li, and Yue Li. 2025.
\newblock A unified solution to diverse heterogeneities in one-shot federated learning.
\newblock In \emph{Proceedings of the 31st ACM SIGKDD Conference on Knowledge Discovery and Data Mining V. 2}, pages 71--82.

\bibitem[{Hong et~al.(2024)Hong, Zhuge, Chen, Zheng, Cheng, Wang, Zhang, Wang, Yau, Lin, Zhou, Ran, Xiao, Wu, and Schmidhuber}]{hong2024metagpt}
Sirui Hong, Mingchen Zhuge, Jonathan Chen, Xiawu Zheng, Yuheng Cheng, Jinlin Wang, Ceyao Zhang, Zili Wang, Steven Ka~Shing Yau, Zijuan Lin, Liyang Zhou, Chenyu Ran, Lingfeng Xiao, Chenglin Wu, and J{\"u}rgen Schmidhuber. 2024.
\newblock \href {https://openreview.net/forum?id=VtmBAGCN7o} {Meta{GPT}: Meta programming for a multi-agent collaborative framework}.
\newblock In \emph{The Twelfth International Conference on Learning Representations}.

\bibitem[{Hu et~al.(2026)Hu, Zhou, Fan, Nie, Ye, Xia, Sun, Jin, Li, Zhang et~al.}]{hu2026owl}
Mengkang Hu, Yuhang Zhou, Wendong Fan, Yuzhou Nie, Ziyu Ye, Bowei Xia, Tao Sun, Zhaoxuan Jin, Yingru Li, Zeyu Zhang, and 1 others. 2026.
\newblock Owl: Optimized workforce learning for general multi-agent assistance in real-world task automation.
\newblock \emph{Advances in Neural Information Processing Systems}, 38:50859--50906.

\bibitem[{Jimenez et~al.(2024)Jimenez, Yang, Wettig, Yao, Pei, Press, and Narasimhan}]{jimenez2024swe}
Carlos~E Jimenez, John Yang, Alexander Wettig, Shunyu Yao, Kexin Pei, Ofir Press, and Karthik Narasimhan. 2024.
\newblock Swe-bench: Can language models resolve real-world github issues?
\newblock In \emph{International Conference on Learning Representations}, volume 2024, pages 54107--54157.

\bibitem[{McMahan et~al.(2017)McMahan, Moore, Ramage, Hampson, and y~Arcas}]{mcmahan2017communication}
Brendan McMahan, Eider Moore, Daniel Ramage, Seth Hampson, and Blaise~Aguera y~Arcas. 2017.
\newblock Communication-efficient learning of deep networks from decentralized data.
\newblock In \emph{Artificial intelligence and statistics}, pages 1273--1282. Pmlr.

\bibitem[{Mialon et~al.(2024)Mialon, Fourrier, Wolf, LeCun, and Scialom}]{mialon2024gaia}
Gr{\'e}goire Mialon, Cl{\'e}mentine Fourrier, Thomas Wolf, Yann LeCun, and Thomas Scialom. 2024.
\newblock Gaia: a benchmark for general ai assistants.
\newblock In \emph{International Conference on Learning Representations}, volume 2024, pages 9025--9049.

\bibitem[{Qin et~al.(2024)Qin, Liang, Ye, Zhu, Yan, Lu, Lin, Cong, Tang, Qian et~al.}]{qin2024toolllm}
Yujia Qin, Shihao Liang, Yining Ye, Kunlun Zhu, Lan Yan, Yaxi Lu, Yankai Lin, Xin Cong, Xiangru Tang, Bill Qian, and 1 others. 2024.
\newblock Toolllm: Facilitating large language models to master 16000+ real-world apis.
\newblock In \emph{International Conference on Learning Representations}, volume 2024, pages 9695--9717.

\bibitem[{Roucher et~al.(2025)Roucher, del Moral, Wolf, von Werra, and Kaunismäki}]{smolagents}
Aymeric Roucher, Albert~Villanova del Moral, Thomas Wolf, Leandro von Werra, and Erik Kaunismäki. 2025.
\newblock `smolagents`: a smol library to build great agentic systems.
\newblock \url{https://github.com/huggingface/smolagents}.

\bibitem[{Shinn et~al.(2023)Shinn, Cassano, Gopinath, Narasimhan, and Yao}]{shinn2023reflexion}
Noah Shinn, Federico Cassano, Ashwin Gopinath, Karthik Narasimhan, and Shunyu Yao. 2023.
\newblock Reflexion: Language agents with verbal reinforcement learning.
\newblock \emph{Advances in neural information processing systems}, 36:8634--8652.

\bibitem[{Tang et~al.(2025)Tang, Qin, Peng, Zhou, Shao, Du, Wei, Zhu, Zhang, Liu, Wang, Hong, Wu, and Zhou}]{tang2025agentkb}
Xiangru Tang, Tianrui Qin, Tianhao Peng, Ziyang Zhou, Yanjun Shao, Tingting Du, Xinming Wei, He~Zhu, Ge~Zhang, Jiaheng Liu, Xingyao Wang, Sirui Hong, Chenglin Wu, and Wangchunshu Zhou. 2025.
\newblock \href {https://openreview.net/forum?id=ohXoWHlrn8} {{AGENT} {KB}: A hierarchical memory framework for cross-domain agentic problem solving}.
\newblock In \emph{ICML 2025 Workshop on Collaborative and Federated Agentic Workflows}.

\bibitem[{Wang et~al.(2025)Wang, Li, Song, Xu, Tang, Zhuge, Pan, Song, Li, Singh et~al.}]{wang2025openhands}
Xingyao Wang, Boxuan Li, Yufan Song, Frank~F Xu, Xiangru Tang, Mingchen Zhuge, Jiayi Pan, Yueqi Song, Bowen Li, Jaskirat Singh, and 1 others. 2025.
\newblock Openhands: An open platform for ai software developers as generalist agents.
\newblock In \emph{International Conference on Learning Representations}, volume 2025, pages 65882--65919.

\bibitem[{Wang et~al.(2024)Wang, Mao, Fried, and Neubig}]{wang2024agent}
Zora~Zhiruo Wang, Jiayuan Mao, Daniel Fried, and Graham Neubig. 2024.
\newblock Agent workflow memory.
\newblock \emph{arXiv preprint arXiv:2409.07429}.

\bibitem[{Xu et~al.(2026{\natexlab{a}})Xu, Koesdwiady, Bei, Han, Huang, Wang, Chen, Wang, Wang, Li et~al.}]{xu2026rethinking}
Jiawei Xu, Arief Koesdwiady, Sisong Bei, Yan Han, Baixiang Huang, Dakuo Wang, Yutong Chen, Zheshen Wang, Peihao Wang, Pan Li, and 1 others. 2026{\natexlab{a}}.
\newblock Rethinking the value of multi-agent workflow: A strong single agent baseline.
\newblock \emph{arXiv preprint arXiv:2601.12307}.

\bibitem[{Xu et~al.(2026{\natexlab{b}})Xu, Liang, Mei, Gao, Tan, and Zhang}]{xu2026mem}
Wujiang Xu, Zujie Liang, Kai Mei, Hang Gao, Juntao Tan, and Yongfeng Zhang. 2026{\natexlab{b}}.
\newblock A-mem: Agentic memory for llm agents.
\newblock \emph{Advances in Neural Information Processing Systems}, 38:17577--17604.

\bibitem[{Yang et~al.(2024)Yang, Jimenez, Wettig, Lieret, Yao, Narasimhan, and Press}]{yang2024swe}
John Yang, Carlos~E Jimenez, Alexander Wettig, Kilian Lieret, Shunyu Yao, Karthik Narasimhan, and Ofir Press. 2024.
\newblock Swe-agent: Agent-computer interfaces enable automated software engineering.
\newblock \emph{Advances in Neural Information Processing Systems}, 37:50528--50652.

\bibitem[{Yao et~al.(2022)Yao, Zhao, Yu, Du, Shafran, Narasimhan, and Cao}]{yao2022react}
Shunyu Yao, Jeffrey Zhao, Dian Yu, Nan Du, Izhak Shafran, Karthik Narasimhan, and Yuan Cao. 2022.
\newblock React: Synergizing reasoning and acting in language models.
\newblock \emph{arXiv preprint arXiv:2210.03629}.

\bibitem[{Zhu et~al.(2025)Zhu, Qiao, Ou, Deng, Lyu, Shen, Liang, Gu, Chen, and Zhang}]{zhu2025knowagent}
Yuqi Zhu, Shuofei Qiao, Yixin Ou, Shumin Deng, Shiwei Lyu, Yue Shen, Lei Liang, Jinjie Gu, Huajun Chen, and Ningyu Zhang. 2025.
\newblock Knowagent: Knowledge-augmented planning for llm-based agents.
\newblock In \emph{Findings of the Association for Computational Linguistics: NAACL 2025}, pages 3709--3732.

\end{thebibliography}

\appendix

\section{FedAgentKE Algorithm}
\label{app:algorithm}

\begin{algorithm}[!h]
\caption{FedAgentKE: Federated Semantic Knowledge Evolution}
\label{alg:fedagentke}
\small
\begin{algorithmic}[1]

\Require Heterogeneous agent clients $\mathcal C=\{C_i\}_{i=1}^{N}$, communication rounds $T$, initial knowledge pool $K_g^{(0)}$

\For{$t=0,\ldots,T-1$}

    \State Server distributes federated knowledge pool $K_g^{(t)}$

    \For{each client $C_i$ in parallel}

        \State Retrieve top-$k$ federated knowledge \\ \hfill
        $\hat K_i^{(t)}=\operatorname{TopK}_{z\in K_g^{(t)}} \operatorname{sim}(\psi(x_i),\phi(z))$

        \State Adapt federated knowledge \\ \hfill
        $\tilde K_i^{(t)}=\mathcal T_i(\hat K_i^{(t)},c_i)$

        \State Refine local reasoning \\ \hfill
        $\rho_i'=\mathcal R_i(\rho_i,\tilde K_i^{(t)})$

        \State Execute local tasks and collect trajectories \\ \hfill
        $\tau_i^{(t)}=\{(x,a^{1:T},o^{1:T},y)\}$

        \State Distill trajectories into knowledge units \\ \hfill
        $K_i^{(t)}=\mathcal D_i(\tau_i^{(t)})$

        \State Upload knowledge units $K_i^{(t)}$

    \EndFor

    \State Construct uploaded knowledge pool \\ \hfill
    $\mathcal U^{(t)}=\bigcup_{i=1}^{N} K_i^{(t)}$

    \State Encode knowledge representations $h_z=\phi(z)$

    \State Perform cross-task semantic clustering \\ \hfill
    $\{\mathcal G_1,\ldots,\mathcal G_M\}=\operatorname{Cluster}(\{h_z\})$

    \For{each cluster $\mathcal G_m$}

        \State Select representative knowledge \\ \hfill
        $\bar z_m=\arg\max_{z\in\mathcal G_m} [\lambda_1 u(z)+\lambda_2 q(z)]$

    \EndFor

    \State Update federated knowledge pool \\ \hfill
    $K_g^{(t+1)}=\operatorname{Update}(K_g^{(t)},\{\bar z_1,\ldots,\bar z_M\})$

\EndFor

\State \Return Final federated knowledge pool $K_g^{(T)}$

\end{algorithmic}
\end{algorithm}

\section{Utility-based Interpretation}
\label{app:theory}

FedAgentKE is motivated by the observation that many reasoning strategies and execution experiences remain transferable across heterogeneous agents despite differences in reasoning protocols, tool environments, and execution workflows.

Let $\rho_i$ denote the local reasoning workflow of client $C_i$. Given adapted federated knowledge $\tilde K_i$, FedAgentKE assumes that reasoning refinement can improve local execution quality:
\begin{equation}
\rho_i'
=
\mathcal R_i(\rho_i,\tilde K_i),
\end{equation}
where $\mathcal R_i(\cdot)$ denotes the local reasoning refinement process.

The utility of knowledge unit $z$ on client $C_i$ is measured by the resulting performance improvement:
\begin{equation}
\Delta_i(z)
=
Q_i(\rho_i')
-
Q_i(\rho_i),
\end{equation}
where $Q_i(\cdot)$ denotes the local execution quality, such as task success rate, reasoning correctness, or execution efficiency.

A knowledge unit is considered transferable if it consistently improves performance across heterogeneous agents:
\begin{equation}
u(z)
=
\frac{1}{N}
\sum_{i=1}^{N}
\Delta_i(z).
\end{equation}

This interpretation motivates the utility-aware representative selection mechanism in FedAgentKE, where knowledge with stronger cross-agent transferability is preferentially preserved during federated aggregation.

Furthermore, semantically redundant reasoning abstractions may reduce knowledge diversity and collaborative transferability across heterogeneous tasks. FedAgentKE therefore performs semantic clustering and redundancy filtering to preserve compact and transferable federated knowledge.

\section{Federated Semantic Knowledge Evolution}
\label{app:semantic_update}

FedAgentKE maintains a global federated knowledge pool that continuously evolves through iterative synchronization across heterogeneous agents. Given representative knowledge units
$
\{\bar z_1,\ldots,\bar z_M\},
$
selected from the uploaded knowledge pool at communication round $t$, the server updates the federated knowledge pool through transferability-aware semantic evolution.

\subsection{Semantic Similarity and Utility Evaluation}

To reduce redundancy, FedAgentKE first identifies semantically overlapping knowledge units based on embedding similarity:
\begin{equation}
s(z_a,z_b)
=
\frac{
\phi(z_a)^\top \phi(z_b)
}{
\|\phi(z_a)\| \|\phi(z_b)\|
},
\end{equation}
where $\phi(\cdot)$ denotes the semantic embedding function.

A newly uploaded knowledge unit $\bar z_m$ is considered redundant to an existing global knowledge unit $z \in K_g^{(t)}$ if:
\begin{equation}
s(\bar z_m,z)
>
\delta,
\end{equation}
where $\delta$ denotes the semantic similarity threshold.

The utility score stored in each knowledge unit is dynamically updated through cross-agent execution feedback across communication rounds. Let $u_z^{(t)}$ denote the utility score of knowledge unit $z$ at communication round $t$. FedAgentKE collects cross-agent utility feedback:
\begin{equation}
r_z^{(t)}
=
\frac{1}{|\mathcal C_z|}
\sum_{C_i \in \mathcal C_z}
\Delta_i(z),
\end{equation}
where $\mathcal C_z$ denotes the set of agents that utilized knowledge unit $z$, and $\Delta_i(z)$ measures the observed execution improvement after applying $z$ on client $C_i$.

The utility score is updated through exponential utility averaging:
\begin{equation}
u_z^{(t+1)}
=
(1-\eta)u_z^{(t)}
+
\eta r_z^{(t)},
\end{equation}
where $\eta$ denotes the utility update rate.

\subsection{Transferability-aware Knowledge Evolution}

When a newly uploaded knowledge unit $\bar z_m$ is semantically similar to an existing global knowledge unit $z$, FedAgentKE preferentially preserves the knowledge unit with stronger cross-agent utility:
\begin{equation}
K_g^{(t+1)}
=
(K_g^{(t)} \setminus \{z\})
\cup
\{\bar z_m\},
\quad
\text{if }
u_{\bar z_m}^{(t+1)}
>
u_z^{(t+1)}.
\end{equation}

Otherwise, the original knowledge unit is retained. This mechanism allows low-utility reasoning abstractions to gradually be replaced by more transferable knowledge discovered from heterogeneous agents.

The global federated knowledge pool is continuously updated through semantic deduplication, utility-aware replacement, and representative knowledge integration:
\begin{equation}
K_g^{(t+1)}
=
\operatorname{Update}
\left(
K_g^{(t)},
\{\bar z_1,\ldots,\bar z_M\}
\right).
\end{equation}

As communication rounds continue, the federated knowledge pool gradually evolves toward more transferable reasoning abstractions across heterogeneous agents and tasks.

\section{Process-level Case Study}
\label{app:process_case}

This appendix presents a process-level case study from a GAIA spreadsheet task. The example illustrates how an initially incorrect execution is converted into reusable semantic knowledge, synchronized through the federated knowledge pool, and later reused in a replay-style diagnostic setting. The reference answer is used only by the benchmark grader and is not inserted into the agent prompt.

\subsection{Task and Initial Failure}

FedAgentKE first records a local execution trajectory
$
\tau_{i,j}=(x_{i,j},a^{1:T}_{i,j},o^{1:T}_{i,j},y_{i,j}),
$
where $x_{i,j}$ is the model-facing task prompt, $a^{1:T}_{i,j}$ and $o^{1:T}_{i,j}$ are the action-observation sequence, and $y_{i,j}$ is the benchmark outcome. In this trajectory, file inspection, code execution, and web search are treated as tool actions inside $a^{1:T}_{i,j}$, with their returned contents recorded as observations in $o^{1:T}_{i,j}$.

\begin{fedagentbox}{Task Input}
\small
\textbf{Run metadata.}
Benchmark: GAIA Level 3. Client framework: SmolAgents. These metadata fields are recorded by the FedAgentKE runner and are not inserted as natural-language task instructions.

\textbf{Agent instruction.}
You have one question to answer. It is paramount that you provide a correct answer. Give it all you can: you have access to the relevant tools to solve it and find the correct answer. Run verification steps if needed, and make sure you find the correct answer.

\textbf{Question.}
The attached spreadsheet contains a list of books I read in the year 2022. What is the title of the book that I read the slowest, using the rate of words per day?

\textbf{Attached file preview.}
The spreadsheet contains columns for title, author, genre, start date, end date, and rating. It does not directly provide word counts.
\end{fedagentbox}

Given this prompt, the client begins by inspecting the attached spreadsheet. The following abridged trajectory shows the key failure: the agent correctly identifies that the reading rate requires both reading duration and book length, but it does not reliably ground the missing word-count evidence before computing the final answer.

\begin{fedagentbox}{Abridged Raw Trajectory Log}
\small
\textbf{Planning.}
The agent identifies the task as an attached-file reasoning problem. It plans to read the spreadsheet, extract each book title and reading interval, obtain or infer word counts, and compute words per day.

\textbf{Action.}
Inspect the attached spreadsheet as text.

\textbf{Observation.}
The spreadsheet lists book titles, authors, genres, start dates, end dates, and ratings. It includes entries such as \textit{Fire and Blood}, \textit{Song of Solomon}, \textit{Out of the Silent Planet}, and \textit{The Shining}, but it does not include word counts.

\textbf{Action.}
Compute reading durations from the spreadsheet dates and fill in approximate word-count estimates for the listed books.

\textbf{Observation.}
The calculation depends on unstable or approximate word-count values rather than verified evidence for each title.

\textbf{Final response.}
Fire and Blood

\textbf{Grading outcome.}
Incorrect.
\end{fedagentbox}

\subsection{Failure-to-Knowledge Distillation}

The failed trajectory is still useful because it exposes a procedural gap: the agent inspected the spreadsheet but did not reliably ground the missing word-count evidence before computing the reading rate. FedAgentKE therefore distills the trajectory into a semantic knowledge unit rather than storing the raw log or treating the wrong answer as factual evidence.

For readability, we present the semantic knowledge unit using descriptive field names. These fields correspond to the tuple
$
z_{i,j}=(\pi_{i,j},\rho_{i,j},e_{i,j},c_{i,j},u_{i,j})
$
defined in Eq.~4: task abstraction $\pi_{i,j}$, reusable reasoning workflow $\rho_{i,j}$, execution correction or failure pattern $e_{i,j}$, framework context $c_{i,j}$, and utility signal $u_{i,j}$.

\begin{fedagentbox}{Failure-derived Semantic Knowledge Unit}
\small
\textbf{Task abstraction .}
Spreadsheet reasoning task requiring a derived numerical comparison. The attached file provides book titles and reading dates, but not all quantities needed for the final words-per-day calculation.

\textbf{Reusable reasoning workflow.}
Inspect the spreadsheet to extract every title and reading interval. Compute reading duration consistently for every row. Check whether the spreadsheet provides word counts. If word counts are missing, retrieve them from reliable external sources, match each value back to the correct spreadsheet row, compute words per day for every title, and select the minimum rate.

\textbf{Failure / correction pattern.}
The failed run used approximate or unstable word-count estimates. Future runs should treat word count as a missing variable to be grounded, not as background knowledge to guess from memory.

\textbf{Framework context.}
The trajectory comes from a GAIA spreadsheet task solved by a SmolAgents-style tool-using client. The stored knowledge abstracts away the client-specific tool interface.

\textbf{Utility signal.}
The trajectory is marked as failure-derived knowledge. Its utility is procedural: it warns later clients to verify missing evidence before computing the final comparison.
\end{fedagentbox}

\subsection{Federated Aggregation}

After local distillation, FedAgentKE synchronizes the semantic knowledge unit with the federated knowledge pool. When later clients contribute related observations, the system can aggregate compatible procedural knowledge. The aggregated unit preserves the original failure signal while adding a more complete reasoning workflow.

\begin{fedagentbox}{Aggregated Semantic Knowledge Unit}
\small
\textbf{Task abstraction.}
Attached-spreadsheet reasoning task where the answer depends on combining file-derived rows with an externally verified missing variable.

\textbf{Reusable reasoning workflow.}
First inspect the spreadsheet and extract titles, dates, and row-level constraints. Then identify which variable required by the question is absent from the file. Retrieve that missing variable from reliable sources, map each retrieved value back to the correct row, compute the requested derived quantity consistently, and verify the selected minimum before answering.

\textbf{Failure / correction pattern.}
Do not replace missing evidence with rough estimates. Keep a traceable mapping between spreadsheet rows, external evidence, derived calculations, and the final selected answer.

\textbf{Framework context.}
The knowledge is stored as framework-agnostic procedural guidance so that later clients may reuse it even if their tool APIs or prompting format differ.

\textbf{Utility signal.}
The unit is useful when a later task requires structured file inspection plus external evidence grounding before a numerical comparison.
\end{fedagentbox}

\subsection{Later Correct Execution}

In a replay-style process analysis, a later client is evaluated on the same benchmark instance after the federated knowledge pool has been updated. This diagnostic setting is used to inspect how the memory changes the reasoning process; the reference answer is not injected into the prompt, and the retrieved knowledge is procedural rather than an answer cache.

\begin{fedagentbox}{Later Run with Adapted Federated Knowledge}
\small
\textbf{Retrieved guidance.}
Treat word count as a missing variable. Inspect the spreadsheet for titles and dates, retrieve reliable word-count evidence, compute words per day for every title, and recheck the minimum-rate candidate.

\textbf{Corrected reasoning pattern.}
The later client separates the task into two evidence sources: the spreadsheet provides reading intervals, while external retrieval provides word counts. The client then computes the words-per-day rate for each book under a consistent formula.

\textbf{Later prediction.}
Out of the Silent Planet

\textbf{Grading outcome.}
Correct.
\end{fedagentbox}

\subsection{Interpretation}

This case study illustrates the intended role of FedAgentKE in a controlled replay setting. The federated knowledge pool is not used as an answer cache. The useful content is the procedural knowledge that the missing word-count variable must be grounded before computing the rate. The initial failed trajectory contributes a warning about an unreliable shortcut, while later aggregation preserves a more complete workflow for future clients.

What transfers is the semantic knowledge unit: inspect the structured file, identify missing variables, retrieve only the missing evidence, compute the derived quantity consistently, and verify the final answer format. This matches the formulation in Eq.~4, where a knowledge unit stores task abstraction, reasoning workflow, correction pattern, framework context, and utility signal rather than raw trajectories or benchmark answers.

\end{document}